\documentclass[draftclsnofoot]{IEEEtran}
\usepackage{graphicx}
\usepackage{epstopdf}

\usepackage{subfigure}
\usepackage{booktabs}
\usepackage{caption}
\usepackage{color}
\usepackage{bm}
\usepackage{CJK}
\usepackage{indentfirst}
\usepackage{amsmath}
\usepackage{amssymb}
\usepackage{float}
\usepackage{multirow}
\usepackage{algorithm}
\usepackage{algpseudocode}
\usepackage{amsmath}
\usepackage{flushend}
\usepackage{geometry}
\newtheorem{proposition}{Proposition}

\begin{document}
\title{Sparse Array of Sub-surface Aided Anti-blockage mmWave Communication Systems}

\author{\IEEEauthorblockN{Weicong~Chen, Xi~Yang,
~Shi~Jin, and Pingping Xu}\\
\thanks{{W. Chen, X. Yang, S. Jin, and P. Xu are with the National Mobile Communications Research Laboratory, Southeast University,
Nanjing, 210096, P. R. China (e-mail: cwc@seu.edu.cn; ouyangxi@seu.edu.cn; jinshi@seu.edu.cn; xpp@seu.edu.cn).} }}

\maketitle
\pagestyle{empty}  
\thispagestyle{empty} 
\begin{abstract}
Recently, reconfigurable intelligent surfaces (RISs) have drawn intensive attention to enhance the coverage of millimeter wave (mmWave) communication systems. However, existing works mainly consider the RIS as a whole uniform plane, which may be unrealistic to be installed on the facade of buildings when the RIS is extreme large. To address this problem, in this paper, we propose a sparse array of sub-surface (SAoS) architecture for RIS, which contains several  rectangle shaped sub-surfaces termed as RIS tiles that can be sparsely deployed. An approximated ergodic spectral efficiency of the SAoS aided system is derived and the performance impact of the SAoS design is evaluated. Based on the approximated ergodic spectral efficiency, we obtain an optimal reflection coefficient design for each RIS tile. Analytical results show that the received signal-to-noise ratios can grow quadratically and linearly to the number of RIS elements under strong and weak LoS scenarios, respectively. Furthermore, we consider the visible region (VR) phenomenon in the SAoS aided mmWave system and find that the optimal distance between RIS tiles is supposed to yield a total SAoS VR  nearly covering the whole blind coverage area. The numerical results verify the tightness of the approximated ergodic spectral efficiency and demonstrate the great system performance.
\end{abstract}


\section{Introduction}
As a promising technology for future wireless communication systems, millimeter wave (mmWave) communication can support giga-bit-per-second data rates since there are abundant of bandwidth available at mmWave frequency spectrum \cite{T. Nitsche}. Although the directional transmission property of mmWave massive multiple-input-multiple-output (MIMO) can compensate the severe propagation loss , which is brought by the high operating frequency, this highly directional transmission also makes the mmWave communication link more susceptible to the blockage \cite{M. Polese}\cite{multi-beam}. Therefore, how to enhance the coverage and improve the anti-blockage ability of mmWave communication systems, is an important issue that is worth studying. \par


Reconfigurable intelligent surfaces (RISs) are an emerging technology with promising potential of constructing smart radio environment \cite{smart-radio}. By exploring the properties of meta-surface, RISs can manipulate the phase of the incident electromagnetic wave and reflect it to the desired direction. Therefore, we can exploit RISs to provide an artificial communication link to users in the blind coverage area. Recently, researches \cite{Y. Han-LIS}-\cite{weighted-sum-rate} focused on RIS are springing up. Considering single-user scenario, \cite{Y. Han-LIS} evaluated the downlink performance of a RIS-assisted large-scale antenna system by formulating a tight upper bound of ergodic spectral efficiency. Aiming at maximizing the energy efficiency in downlink multi-user communication, the authors in \cite{C. Huang} proposed two effective approaches to tackle the non-convex optimization problems by means of designing both the transmit power allocation and the RIS phase shifts. By jointly optimizing the active beamforming at the base station (BS) and the passive beamforming at the RIS, the maximal received signal power for a single-antenna user, the maximal sum rate and weighted sum rate for multiple users were investigated in \cite{Q. Wu}, \cite{sum-rate} and \cite{weighted-sum-rate}, respectively. \par
However, the aforementioned works all consider the RIS as a whole uniform plane. In practical, RISs are supposed to be deployed on the facade of the building, which, unfortunately, means that a whole compact large-scale RIS plane will be unrealistic, since the facade of buildings has their own functional requirement such as, daylighting, commercial advertisement display with digital screen, and so on.\par

To solve this problem, in this paper,  a sparse array of sub-surface (SAoS)-based architecture is proposed for the RIS to assist the mmWave communication system. In SAoS, the RIS is divided into several rectangle-shaped sub-surfaces . Each of these sub-surfaces is termed as an RIS tile and can be sparsely deployed. Meanwhile, visible region (VR) is also considered in each RIS tile given by the large effective antenna array aperture. To investigate the benefits of the SAoS\footnote{Note that in the following analysis, we use SAoS to represent the SAoS-based RIS.}, we first derive a tight approximation for the ergodic system spectral efficiency. Then, based on the approximated ergodic spectral efficiency, the optimal reflection coefficients of SAoS are obtained. The analytical results show that the received SNR can grow quadratically to the element number of SAoS when strong line-of-sight (LoS) paths exist between both BS-SAoS channels and mobile station (MS)-SAoS channels. When LoS paths are absent either between BS-SAoS channels or MS-SAoS channels, the received SNR grows linearly to the element number of SAoS. Numerical results validate the tightness of the approximated ergodic spectral efficieny and demonstrate the significant performance gain acquired by the optimal reflection coefficient design. The scaling law under different Rician factor is also investigated. Moreover, it is found that when the distance between RIS tiles is designed to maximize the ergodic system spectral efficiency,  the total VR of the SAoS nearly covers  the whole blind coverage area.\par


\emph{Notations}: The lowercase and uppercase of a letter denote the vector and matrix, respectively. The transpose and conjugated-transpose operation are represented by the superscripts $(.)^T$ and $(.)^H$, respectively. ${\rm tr}(.)$ calculates the trace of a matrix. $|.|$, $\|.\|$ and are used to indicate the absolute value, Euclidean norm, respectively. The kronecker product is denoted by $\otimes$. $\lceil.\rceil$ is the integer ceiling. ${\mathbb E}\{.\}$ is to calculate the mean. ${\rm blkdiag}\{{\bf X}_1,{\bf X}_2,\cdots,{\bf X}_N\}$ represents a diagonal matrix with block diagonal matrices ${\bf X}_i$, $i=1,\cdots,N$, while ${\rm diag}(a_1,a_2,\cdots,a_N)$ indicates a diagonal matrix with diagonal elements $a_i$, $i=1,\cdots,N$.

\section{System model}\label{sec:2}

\begin{figure}[!t]
\centering
    \includegraphics[width=0.5\textwidth]{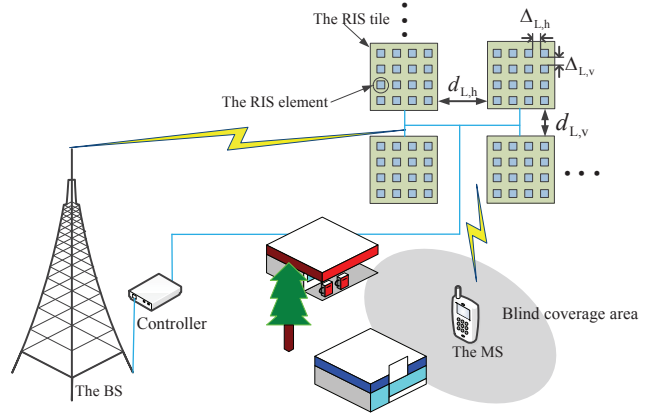}
\caption{Sparse array of sub-surface aided anti-blockage mmWave communication systems}
\label{Fig.system_model}
\vspace{-0.3cm}
\end{figure}

We consider a SAoS-aided mmWave uplink communication system, as shown in Fig. \ref{Fig.system_model}. A single-antenna MS is located in the blind coverage area and cannot directly communicate with the BS. To deal with the blockage problem, the SAoS, acting as an anomalous mirror, is deployed on a two dimensional facade of a building which is in the line of sight of the BS. The BS is equipped with a uniform planar array (UPA) with $N_{\rm B} = N_{\rm B,v}\times N_{\rm B,h}$ antennas, and the vertical and horizontal spacing between adjacent antennas is half carrier wavelength. We define each rectangle-shaped sub-surface as an RIS tile, as shown in Fig.\ref{Fig.system_model}. The SAoS contains $M_{\rm L} = M_{\rm L,v}\times M_{\rm L,h}$ RIS tiles with adjacent vertical and horizontal distance being $d_{\rm L,v}$ and $d_{\rm L,h}$, respectively. For each RIS tile, there are $N_{\rm L} = N_{\rm L,v}\times N_{\rm L,h}$ RIS elements spacing $\Delta_{\rm L,v}$ and $\Delta_{\rm L,h}$ in the vertical and horizontal axis, respectively. The RIS elements are controlled by the BS via an RIS controller and can be equivalently regarded as phase shifters which can manipulate the phase of the incident signal. Note that the SAoS is deployed close to the blocked MS but away from the BS.

\par

\subsection{Channel Model}\label{sec2.2}
Since the distance from the SAoS to the BS is far enough, the widely used Saleh-Valenzuela geometric model \cite{SVchannel}\cite{W. Chen} can be adopted to describe the channel between the BS and RIS tiles. With Rician factor $K_{\rm B}$ denoting the ratio between the power of the LoS component and the non-LoS (NLoS) component, the channel between the BS and the $m$-th RIS tile can be expressed as
\begin{equation}\label{eq:H_m}
  {{\bf{H}}_m} = \sqrt {\frac{{{K_{\rm{B}}}}}{{{K_{\rm{B}}} + 1}}} {{\bf{H}}_{m,{\rm{L}}}} + \sqrt {\frac{1}{{{K_{\rm{B}}} + 1}}} {{\bf{H}}_{m,{\rm{N}}}},
\end{equation}
where the LoS component is given by
\begin{equation}\label{eq:H_m,L}
\begin{aligned}
  {{\bf{H}}_{m,{\rm{L}}}} = {\beta _m}\sqrt {{N_{\rm{B}}}{N_{\rm{L}}}} {{\bf{a}}_{\rm{r}}}\left( {{\Theta _{m{\rm{,r}}}},{\Phi _{m{\rm{,r}}}}} \right){\bf{a}}_{\rm{t}}^H\left( {{\Theta _{m{\rm{,t}}}},{\Phi _{m,{\rm{t}}}}} \right),
  \end{aligned}
\end{equation}
where ${\beta _m}= 1/\sqrt {4\pi R_m^2}$ is the free space loss attenuation \cite{M. Jung}, ${\Theta _{m{\rm{,r}}}} = \pi \cos {\theta _{m{\rm{,r}}}}$ and ${\Phi _{m{\rm{,r}}}} = \pi \sin {\theta _{m{\rm{,r}}}}\sin {\phi _{m{\rm{,r}}}}$ are spatial frequencies at the BS, and ${\theta _{m{\rm{,r}}}}$ and ${\phi _{m{\rm{,r}}}}$ denote the elevation and azimuth angles of arrival (AoA) for LoS path, respectively. ${\Theta _{m{\rm{,t}}}} = \frac{{2\pi {\Delta _{\rm{v}}}}}{\lambda }\cos {\theta _{m{\rm{,t}}}}$ and ${\Phi _{m{\rm{,t}}}} = \frac{{2\pi {\Delta _{\rm{h}}}}}{\lambda }\sin {\theta _{m{\rm{,t}}}}\sin {\phi _{m{\rm{,t}}}}$ are spatial frequencies at the $m$-th RIS tile with the elevation and azimuth angles of departure (AoD) being ${\theta _{m{\rm{,t}}}}$ and ${\phi _{m{\rm{,t}}}}$, respectively. In \eqref{eq:H_m,L}, ${\bf{a}}\left( {\Theta ,\Phi } \right) = {{\bf{a}}_{\rm{v}}}\left( \Theta  \right) \otimes {{\bf{a}}_{\rm{h}}}\left( \Phi  \right)$ denotes the array response vector of UPA, where ${{\bf{a}}_{\rm{v}}}\left( \Theta  \right)$ and $ {{\bf{a}}_{\rm{h}}}\left( \Phi  \right)$ are the vertical and horizontal array response vectors of uniform linear array (ULA), respectively, and can be expressed as
\begin{equation}\label{eq:ARV_ULA}
  {\bf{a}}\left( X \right) = {1}/{{\sqrt N }}{\left[ {\begin{array}{*{20}{c}}
1&{{e^{jX}}}& \cdots &{{e^{j\left( {N - 1} \right)X}}}
\end{array}} \right]^T},
\end{equation}
$X\in\{\Theta,\Phi\}$. The NLoS component of ${\bf H}_m$ is given by ${{\bf{H}}_{m,{\rm{N}}}} = \sqrt {\frac{{{N_{\rm{B}}}{N_{\rm{L}}}}}{{{L_{\rm{B}}}}}} \sum\limits_{l = 1}^{{L_{\rm{B}}}} {{{\bf{H}}_{m,l,{\rm{N}}}}}$,
where ${L_{\rm{B}}}$ is the number of NLoS paths. ${{\bf{H}}_{m,l,{\rm{N}}}}$ is the channel of $l$-th NLoS path
between the BS and the $m$-th RIS tile and can be written as
\begin{equation}\label{eq:H_m,l,N}
\begin{aligned}
  {{\bf{H}}_{m,l,{\rm{N}}}}= & {g_{m,l,{\rm{N}}}}{\beta _{m,l,{\rm{N}}}}{{\bf{a}}_{\rm{r}}}\left( {{\Theta _{m,l,{\rm{N,r}}}},{\Phi _{m,l,{\rm{N,r}}}}} \right)\\
  &\times{\bf{a}}_{\rm{t}}^H\left( {{\Theta _{m,l,{\rm{N,t}}}},{\Phi _{m,l,{\rm{N,t}}}}} \right),
\end{aligned}
\end{equation}
where ${g_{m,l,{\rm{N}}}} \sim {\mathcal {CN}}(0,1)$ and ${\beta _{m,l,{\rm{N}}}}$ are the path gain and the free space loss attenuation of $l$-th NLoS path, respectively.\par
At the MS side, we denote the channel between the MS and the $m$-th RIS tile as
\begin{equation}\label{eq:h_k,m}
  {{\bf{h}}_{m}} = \sqrt {\frac{{{K_{\rm{M}}}}}{{{K_{\rm{M}}} + 1}}} {{\bf{h}}_{m,{\rm{L}}}} + \sqrt {\frac{1}{{{K_{\rm{M}}} + 1}}} {{\bf{h}}_{m,{\rm{N}}}},
\end{equation}
where ${K_{\rm{M}}}$ is the Rician factor. ${{\bf{h}}_{m,{\rm{L}}}}$ is the LoS component of ${{\bf{h}}_{m}}$, whose $c$-th element can be expressed as ${{\bf{h}}_{m,{\rm{L}}}}\left[ c \right] = {\alpha _{m,c}}{e^{j\frac{{2\pi {r_{m,c}}}}{\lambda }}}$,
where ${r_{m,c}}$ and ${\alpha _{m,c}}$ are the LoS distance and the  probabilistic free space loss attenuation between the MS and the $c$-th element of the $m$-th RIS tile, respectively. Since the MS is rather close to the SAoS, it may not see all those RIS tiles. Therefore, we assume each RIS tile has its own VR corresponding to the MS. When the MS is in the VR of the $m$-th RIS tile, ${\alpha _{m,c}}=1/\sqrt {4\pi {r_{m,c}}^2}$, otherwise ${\alpha _{m,c}}=0$. The NLoS component of ${{\bf{h}}_{m}}$ is given by
\begin{equation}\label{eq:h,k,m,N}
  {{\bf{h}}_{m,{\rm{N}}}} = \sqrt {{{{N_{\rm{L}}}}}/{{{L_{m,{\rm{N}}}}}}} \sum\limits_{l = 1}^{{L_{m,{\rm{N}}}}} {{{\bf{h}}_{m,l,{\rm{N}}}}},
\end{equation}
where ${L_{m,{\rm{N}}}}$ is the number of NLoS paths, and ${{\bf{h}}_{m,l,{\rm{N}}}}$, the $l$-th component of ${{\bf{h}}_{m,{\rm{N}}}}$, can be expressed by ${{\bf{h}}_{m,l,{\rm{N}}}} = {g_{m,l,{\rm{N}}}}{\alpha _{m,l,{\rm{N}}}}{\bf{a}}\left( {{\Xi _{m,l}},{\Psi _{m,l}}} \right)$,
where ${g_{m,l,{\rm{N}}}}\sim {\mathcal {CN}}(0,1)$ and ${\alpha _{m,l,{\rm{N}}}}$ are the path gain and the free space loss attenuation, respectively. The spatial frequencies ${\Xi _{m,l}} = \frac{{2\pi {\Delta _{\rm{v}}}}}{\lambda }\cos {\xi _{m,l}}$ and ${\Psi _{m,l}}=\frac{{2\pi {\Delta _{\rm{h}}}}}{\lambda }\sin {\xi _{m,l}}\sin {\psi _{m,l}}$ are expressed in term of AoA $\xi _{m,l}$ and ${\psi _{m,l}}$ of $l$-th NLoS path from the MS to the $m$-th RIS tile.

\subsection{Uplink Ergodic Spectral Efficiency}
In the uplink transmission of SAoS aided anti-blockage mmWave communication system, the received signal at the BS is combined as
\begin{equation}\label{eq:received_sig}
  y = {{\bf w}^H}{{\bf{H}}_{{\rm{BS}}}}{{\bf \Gamma}} {{\bf{h}}}{{x}} + {{\bf w}^H}{\bf{z}},
\end{equation}
where ${{\bf{H}}_{{\rm{BS}}}} = \left[ {{{\bf{H}}_1}, \cdots ,{{\bf{H}}_{{M_{\rm{L}}}}}} \right] \in {\mathbb C}{^{{N_{\rm{B}}} \times {N_{\rm{L}}}{M_{\rm{L}}}}}$ is the channel between the SAoS and the BS, ${{\bf{h}}} = {\left[ {{\bf{h}}_{1}^T, \cdots ,{\bf{h}}_{{M_{\rm{L}}}}^T} \right]^T} \in {\mathbb C} {^{{N_{\rm{L}}}{M_{\rm{L}}} \times 1}}$ is the channel between the MS and the SAoS, ${\bf{\Gamma }} = {\rm blkdiag}\left\{ {{{\bf{\Gamma }}_1}, \cdots ,{{\bf{\Gamma }}_{{M_{\rm{L}}}}}} \right\} \in {\mathbb C} {^{{N_{\rm{L}}}{M_{\rm{L}}} \times {N_{\rm{L}}}{M_{\rm{L}}}}}$ where ${{\bf{\Gamma }}_m} = {\rm diag}( {{e^{j{\varpi _{m,1}}}},,{e^{j{\varpi _{m,{N_{\rm{L}}}}}}}} )$ is the reflect coefficient matrix of the $m$-th RIS tile, $x$ is the transmitted signal of MS satisfying ${\mathbb E}\{|x|^2\}=1$, ${\bf{z}} \in {\mathbb C}{^{{N_{\rm{B}}}}}$ is the complex Gaussian noise satisfying ${\bf{z}}\sim {\mathcal {CN}}(0,{{\sigma ^2}}{\bf I})$ where ${\sigma ^2}$ is the noise power. We consider the BS adopts maximum ratio combining (MRC) receiver, i.e., ${{\bf{w}}} = {{{{\bf{H}}_{{\rm{BS}}}}\Gamma {{\bf{h}}_k}}}/{{\left\| {{{\bf{H}}_{{\rm{BS}}}}\Gamma {{\bf{h}}}} \right\|}}$, then the system ergodic spectral efficiency can be expressed as
\begin{equation}\label{eq:C_single}
C = {\mathbb E}\left\{ {{{\log }_2}\left( {1 + {{\left\| {{{\bf{H}}_{{\rm{BS}}}}{\bf \Gamma} {\bf{h}}} \right\|^2}}/{\left\|{{\bf w}^H}{\bf{z}}\right\|^2}} \right)} \right\}.
\end{equation}\par
Observing at \eqref{eq:C_single}, on one hand, both ${\bf H}_{\rm BS}$, $\bf h$ and ${\bf \Gamma}$ play an important role in the system ergodic spectral efficiency $C$.  On the other hand, from \eqref{eq:H_m,L}, \eqref{eq:H_m,l,N}-\eqref{eq:h,k,m,N}, it indicates that  the distance between each RIS tiles not only effects ${{\bf{H}}_{{\rm{BS}}}}$ and {\bf{h}}, but also determines the total VR of the SAoS. Furthermore, the reflect coefficient matrix ${\bf \Gamma}$ of the SAoS in \eqref{eq:C_single} is also configurable to improve the ergodic spectral efficiency. Therefore, in the following, we devote to investigate the design of SAoS to maximize the above ergodic spectral efficiency.


\section{Ergodic spectral efficiency analysis}\label{sec:3}
In this section, a closed-form approximation is derived for the ergodic spectral efficiency. Based on the approximation, effects brought by the reflection coefficients, deployed position, and distance between RIS tiles of the SAoS are investigated.

\subsection{Approximation of Ergodic Spectral Efficiency}

The approximation of the ergodic spectral efficiency is provided in the following proposition.
\begin{proposition}\label{pro:1}
The ergodic spectral efficiency of sparse array of the sub-surface aided anti-blockage mmWave communication system is approximated by
\begin{equation}\label{eq:C_app_close}
  C_{\rm app}={\log _2}\left( {1 + {S}/{\sigma ^2}} \right),
\end{equation}
where ${\sigma ^2}$ is the noise power,
\begin{equation}\label{eq:S}
  \begin{aligned}
S &= T_1\sum\limits_{m = 1}^{{M_{\rm{L}}}} {\sum\limits_{n = 1}^{{M_{\rm{L}}}} {{\beta _m}{\beta _n}{V_{m,n}}\sum\limits_{c=1}^{{N_{\rm{L}}}} {\sum\limits_{s=1}^{{N_{\rm{L}}}} {{\alpha _{n,c}}{\alpha _{m,s}}{e^{j{\Omega _{n,c,m,s}}}}} } } } \\
& + T{K_{\rm{M}}}{N_{\rm{B}}}\sum\limits_{m = 1}^{{M_{\rm{L}}}} {\sum\limits_{l = 1}^{{L_{\rm{B}}}} {\frac{{\beta _{m,l,{\rm{N}}}^2}}{{{L_{\rm{B}}}}}} \sum\limits_{t = 1}^{{N_{\rm{L}}}} {\alpha _{m,t}^2} } \\
& + T{N_{\rm{B}}}{N_{\rm{L}}}{K_{\rm B}} \sum\limits_{m = 1}^{{M_{\rm{L}}}} {\beta _m^2\sum\limits_{t = 1}^{{L_{m,{\rm{N}}}}} {\frac{{\alpha _{m,t,{\rm{N}}}^2}}{{{L_{m,{\rm{N}}}}}}} } \\
& +T{N_{\rm{B}}}{N_{\rm{L}}}\sum\limits_{m = 1}^{{M_{\rm{L}}}} {\sum\limits_{l = 1}^{{L_{\rm{B}}}} {\frac{{\beta _{m,l,{\rm{N}}}^2}}{{{L_{\rm{B}}}}}} \sum\limits_{t = 1}^{{L_{m,{\rm{N}}}}} {\frac{{\alpha _{m,t,{\rm{N}}}^2}}{{{L_{m,{\rm{N}}}}}}} },
\end{aligned}
\end{equation}
where $T=({{{K_{\rm{M}}} + 1}})^{-1}({{{K_{\rm{B}}} + 1}})^{-1}$, $T_1=T{K_{\rm{M}}}{K_{\rm{B}}}$, $V_{m,n}$ and ${\Omega _{n,c,m,s}}$ are given in \eqref{eq:V_m,n} and \eqref{eq:Omega_k,k}, respectively.
\end{proposition}

\begin{IEEEproof}
Since the first term in \eqref{eq:S} is controllable as described later, we only prove the first term here and omit the proof of the other terms due to the limited space\footnote{In fact, the derivation process of the last three terms in \eqref{eq:S} is similar to the first one, thus we omit the derivation process due to the limited space.}.\par
Utilizes the approximation ${\mathbb E}\left\{ {{{\log }_2}\left( {1 + {x \mathord{\left/
{\vphantom {x y}} \right.
\kern-\nulldelimiterspace} y}} \right)} \right\} \approx {\log _2}\left( {1 + {{{\mathbb E}\left\{ x \right\}} \mathord{\left/
{\vphantom {{\left\{ x \right\}} {\left\{ y \right\}}}} \right.
\kern-\nulldelimiterspace} {{\mathbb E}\left\{ y \right\}}}} \right)$ from \cite{Ck_appro_1}\cite{Ck_appro_2}, the ergodic spectral efficiency can be approximated as
\begin{equation}\label{eq:C_app}
  {C}  \approx C_{\rm app} = {\log _2}\left( {1 + {{{\mathbb E}\left\{ {{\left\| {{{\bf{H}}_{{\rm{BS}}}}{{\bf \Gamma}} {\bf{h}}} \right\|^2}} \right\}}}/{{{\mathbb E}\left\{ {\left\|{{\bf w}^H}{\bf{z}}\right\|^2} \right\}}}} \right)
\end{equation}
In \eqref{eq:C_app}, it is straightforward to have
\begin{equation}\label{eq:noise_power}
\begin{aligned}
  {{\mathbb E}\left\{ {\left\|{{\bf w}^H}{\bf{z}}\right\|^2} \right\}}&={{\mathbb E}\left\{ {{{\bf w}^H}{\bf{z}}} {\bf{z}}^H{\bf w} \right\}}\\
  &=\sigma ^2{{\mathbb E}\left\{ {{{\bf w}^H}} {\bf w} \right\}}=\sigma ^2
  \end{aligned}
\end{equation}
By decomposing ${{{\bf{H}}_{{\rm{BS}}}}{{\bf \Gamma}} {\bf{h}}}$ into the sum of ${{{\bf{H}}_{m}}{\bf \Gamma}_m {{\bf{h}}_{m}}}$ and replacing ${{\bf{H}}_{m}}$ and $ {{\bf{h}}_{m}}$ with \eqref{eq:H_m} and \eqref{eq:h_k,m}, respectively, the mean of received signal power in \eqref{eq:C_app} can be divided into four non-zero parts as\footnote{Due to the limited space, the derivation procedure is omitted in this paper.}
\begin{equation}\label{eq:E_four}
  \begin{aligned}
&{{\mathbb E}\left\{ {{\left\| {{{\bf{H}}_{{\rm{BS}}}}{{\bf \Gamma}} {\bf{h}}} \right\|^2}} \right\}} \\
&= T_1\sum\limits_{m = 1}^{{M_{\rm{L}}}} {\sum\limits_{n = 1}^{{M_{\rm{L}}}} {\mathbb E}{\left\{ {{O_{1,1}}} \right\}} }  + T{K_{\rm{M}}}\sum\limits_{m = 1}^{{M_{\rm{L}}}} {\sum\limits_{n = 1}^{{M_{\rm{L}}}} {\mathbb E}{\left\{ {{O_{1,4}}} \right\}} } \\
& + T{K_{\rm{B}}}\sum\limits_{m = 1}^{{M_{\rm{L}}}} {\sum\limits_{n = 1}^{{M_{\rm{L}}}} {\mathbb E}{\left\{ {{O_{4,1}}} \right\}} }  + T\sum\limits_{m = 1}^{{M_{\rm{L}}}} {\sum\limits_{n = 1}^{{M_{\rm{L}}}} {\mathbb E}{\left\{ {{O_{4,4}}} \right\}} },
\end{aligned}
\end{equation}
where
\begin{equation}\label{eq:O_used}
  \begin{aligned}
{O_{1,1}} = {\bf{h}}_{m,{\rm{L}}}^H{\bf{\Gamma }}_m^H{\bf{H}}_{m,{\rm{L}}}^H{{\bf{H}}_{n,{\rm{L}}}}{{\bf{\Gamma }}_n}{{\bf{h}}_{n,{\rm{L}}}},\\
{O_{1,4}} = {\bf{h}}_{m,{\rm{L}}}^H{\bf{\Gamma }}_m^H{\bf{H}}_{m,{\rm{N}}}^H{{\bf{H}}_{n,{\rm{N}}}}{{\bf{\Gamma }}_n}{{\bf{h}}_{n,{\rm{L}}}},\\
{O_{4,1}} = {\bf{h}}_{m,{\rm{N}}}^H{\bf{\Gamma }}_m^H{\bf{H}}_{m,{\rm{L}}}^H{{\bf{H}}_{n,{\rm{L}}}}{{\bf{\Gamma }}_n}{{\bf{h}}_{n,{\rm{N}}}},\\
{O_{4,4}} = {\bf{h}}_{m,{\rm{N}}}^H{\bf{\Gamma }}_m^H{\bf{H}}_{m,{\rm{N}}}^H{{\bf{H}}_{n,{\rm{N}}}}{{\bf{\Gamma }}_n}{{\bf{h}}_{n,{\rm{N}}}}
\end{aligned}
\end{equation}
It is obviously that ${O_{1,1}}$ is a constant determined by the LoS component of channels, thus ${\mathbb E}\left\{ {{O_{1,1}}} \right\} = {O_{1,1}}$. Using the equation ${{\bf{a}}^H}{\bf{b}} = {\rm{tr}}\left( {{\bf{b}}{{\bf{a}}^H}} \right)$, we have ${O_{1,1}} = {\rm{tr}}\left( {{{\bf{\Gamma }}_n}{{\bf{h}}_{n,{\rm{L}}}}{\bf{h}}_{m,{\rm{L}}}^H{\bf{\Gamma }}_m^H{\bf{H}}_{m,{\rm{L}}}^H{{\bf{H}}_{n,{\rm{L}}}}} \right)$. According to the definition of LoS component for the channel between the BS and the RIS tile, the element in row $s$ and column $c$ of matrix ${\bf{H}}_{m,{\rm{L}}}^H{{\bf{H}}_{n,{\rm{L}}}}$ can be calculated as
\begin{equation}\label{eq:H^HH_s,c}
  {\bf{H}}_{m,{\rm{L}}}^H{{\bf{H}}_{n,{\rm{L}}}}\left( {s,c} \right) = {\beta _m}{\beta _n}{V_{m,n}}{e^{j{\Upsilon _{m,n,c,s}}}},
\end{equation}
where $V_{m,n}$ is the Dirichlet kernel and is given by
\begin{equation}\label{eq:V_m,n}
\begin{aligned}
  &{V_{m,n}}\\
   &= \frac{{\sin ( {\frac{{{N_{{\rm{B,v}}}}}}{2} {{\Theta _{n{\rm{,r}}}} - {\Theta _{m{\rm{,r}}}}} )} )}}{{\sin ( {\frac{1}{2}( {{\Theta _{n{\rm{,r}}}} - {\Theta _{m{\rm{,r}}}}})} )}}\frac{{\sin ( {\frac{{{N_{{\rm{B,h}}}}}}{2}( {{\Phi _{n{\rm{,r}}}} - {\Phi _{m{\rm{,r}}}}})} )}}{{\sin ( {\frac{1}{2}( {{\Phi _{n{\rm{,r}}}} - {\Phi _{m{\rm{,r}}}}})} )}}
  \end{aligned}
\end{equation}
The phase term ${\Upsilon _{m,n,c,s}}$ in \eqref{eq:H^HH_s,c} is calculated as
\begin{equation}\label{eq:Upsilon}
  \begin{aligned}
&{\Upsilon _{m,n,c,s}} = ( {{f_{{N_{{\rm{L,h}}}}}}( s ) - 1} ){\Theta _{m{\rm{,t}}}} + ( {{g_{{N_{{\rm{L,h}}}}}}( s ) - 1} ){\Phi _{m{\rm{,t}}}}  \\
 &-\left( {{f_{{N_{{\rm{L,h}}}}}}\left( c \right) - 1} \right){\Theta _{n{\rm{,t}}}} - \left( {{g_{{N_{{\rm{L,h}}}}}}\left( c \right) - 1} \right){\Phi _{n{\rm{,t}}}}\\
 &+ {{( {{N_{{\rm{B,v}}}} - 1} )( {{\Theta _{n{\rm{,r}}}} - {\Theta _{m{\rm{,r}}}}} ) }}/{2}\\
 &+ {{ ( {{N_{{\rm{B,h}}}} - 1} )( {{\Phi _{n{\rm{,r}}}} - {\Phi _{m{\rm{,r}}}}} )}}/{2},
\end{aligned}
\end{equation}
where the function ${f_{{N_{{\rm{L,h}}}}}}\left( x \right)$ takes the row number of $x$ and is defined as
\begin{equation}\label{eq:f_x}
  {f_{{N_{{\rm{L,h}}}}}}\left( x \right) = \left\lceil {{x}/{{{N_{{\rm{L,h}}}}}}} \right\rceil ,
\end{equation}
and the function ${g_{{N_{{\rm{L,h}}}}}}\left( x \right)$ takes the column number of $x$ and is defined as
\begin{equation}\label{eq:g_x}
  {g_{{N_{{\rm{L,h}}}}}}\left( x \right) = x - \left( {\left\lceil {{x}/{{{N_{{\rm{L,h}}}}}}} \right\rceil  - 1} \right){N_{{\rm{L,h}}}}
\end{equation}
According to the definition of reflect coefficient matrix ${{\bf{\Gamma }}_m}$ and the LoS component for the channel between the MS and the RIS tile, we can derive \eqref{eq:Ga_h_h^H_Ga^H},
\begin{figure*}
\begin{equation}\label{eq:Ga_h_h^H_Ga^H}
  {{\bf{\Gamma }}_n}{{\bf{h}}_{n,{\rm{L}}}}{\bf{h}}_{m,{\rm{L}}}^H{\bf{\Gamma }}_m^H = \left[ {\begin{array}{*{20}{c}}
{{\alpha _{n,1}}{\alpha _{m,1}}{e^{j{P_{n,1,m,1}}}}}& \cdots &{{\alpha _{n,1}}{\alpha _{m,{N_{\rm{L}}}}}{e^{j{P_{n,1,m,{N_{\rm{L}}}}}}}}\\
 \vdots & \ddots & \vdots \\
{{\alpha _{n,{N_{\rm{L}}}}}{\alpha _{m,1}}{e^{j{P_{n,{N_{\rm{L}}},m,1}}}}}& \cdots &{{\alpha _{n,{N_{\rm{L}}}}}{\alpha _{m,{N_{\rm{L}}}}}{e^{j{P_{n,{N_{\rm{L}}},m,{N_{\rm{L}}}}}}}}
\end{array}} \right]
\end{equation}
\vspace{-0.9cm}
\end{figure*}
where $P_{n,c,m,s} = {{2\pi \left( {{r_{n,c}} - {r_{m,s}}} \right)}}/{\lambda } + {\varpi _{n,c}} - {\varpi _{m,s}}$. Multiplying \eqref{eq:H^HH_s,c} and \eqref{eq:Ga_h_h^H_Ga^H}, $O_{1,1}$ is calculated as
\begin{equation}\label{eq_O_1,1,F}
\begin{aligned}
{O_{1,1}} &= {\rm{tr}}\left( {{{\bf{\Gamma }}_n}{{\bf{h}}_{n,{\rm{L}}}}{\bf{h}}_{m,{\rm{L}}}^H{\bf{\Gamma }}_m^H{\bf{H}}_{m,{\rm{L}}}^H{{\bf{H}}_{n,{\rm{L}}}}} \right) \\
&={\beta _m}{\beta _n}{V_{m,n}}\sum\limits_c^{{N_{\rm{L}}}} {\sum\limits_s^{{N_{\rm{L}}}} {{\alpha _{n,c}}{\alpha _{m,s}}{e^{j{\Omega _{n,c,m,s}}}}} },
\end{aligned}
\end{equation}
where ${\Omega _{n,c,m,s}}$ can be expressed as
\begin{equation}\label{eq:Omega_k,k}
  {\Omega _{n,c,m,s}} = P_{n,c,m,s} + {\Upsilon _{m,n,c,s}}
\end{equation}
Substituting \eqref{eq_O_1,1,F} into \eqref{eq:E_four}, then the first term of \eqref{eq:S} is obtained.
\end{IEEEproof}\par

According to \emph{Proposition} \ref{pro:1}, when transceiver parameters (transmit SNR $\frac{1}{\sigma^2}$ and receive antenna number $N_{\rm B}$) and the environment parameters (Rician factor $K_{\rm B}$ and $K_{\rm M}$, NLoS path numbers $L_{\rm B}$ and $L_{m,{\rm N}}$, and NLoS free space loss attenuation $\beta_{m,l,{\rm N}}$ and $\alpha_{m,t,{\rm N}}$) are fixed, the ergodic spectral efficiency is determined by the concrete design parameters of the SAoS (the reflection coefficient in $\Omega _{n,c,m,s}$, the number of RIS pathes $M_{\rm L}$, the elements number in each RIS pathes $N_{\rm L}$ and the deployed position of the RIS tiles that determines the free space attenuation $\beta_m$ and $\alpha_{n,c}$). In the following, we investigate how the SAoS design impacts the ergodic spectral efficiency under the condition that the transceiver and environment parameters keep unchanged.\par

\subsection{Effects of the SAoS Design}\label{sec:3-2}

Note that to highlight the insights, the NLoS free space attenuations between the BS/MS and the RIS tile are assumed to be the same, i.e., $\beta_{m,l,{\rm N}}=\beta_{m,{\rm N}}$ for $\forall l\in L_{\rm B}$ and $\alpha_{m,t,{\rm N}}=\alpha_{m,{\rm N}}$ for $\forall t \in L_{m,{\rm N}}$. In addition, the effects on the approximation of the ergodic spectral efficiency is analyzed indirectly by the mean receive signal power, since $C_{\rm app}$ is only determined by $S$.

\subsubsection{The reflection coefficient design}
As we can see from \eqref{eq:S}, the last three components of $S$ are the sum of real numbers and they are constants when the system is designed. Nevertheless, the first component of $S$ has complex number additive. It is straightforward that setting $ {\Omega _{n,c,m,s}}=0$ can maximize the sum operation. To this end, given $n \in \left\{ {1,2, \cdots ,{M_{\rm{L}}}} \right\}$ and $c \in \left\{ {1,2, \cdots ,{N_{\rm{L}}}} \right\}$ to set ${\varpi _{n,c}}$ as reference, the reflect coefficients ${\varpi _{m,s}}$ can be optimally designed as
\begin{equation}\label{eq:reflect-design}
  {\varpi _{m,s}} = {{2\pi \left( {{r_{k,n,c}} - {r_{k,m,s}}} \right)}}/{\lambda } + {\varpi _{n,c}} + {\Upsilon _{m,n,c,s}}
\end{equation}
for $\forall m \in \left\{ {1,2, \cdots ,{M_{\rm{L}}}} \right\}$ and $\forall s \in \left\{ {1,2, \cdots ,{N_{\rm{L}}}} \right\}$. When the SAoS aided anti-blockage mmWave communication systems are deployed, the optimal reflection coefficient design in \eqref{eq:reflect-design} only requires the position of the MS to calculate $ {r_{k,m,s}}$ for $\forall m \in \left\{ {1,2, \cdots ,{M_{\rm{L}}}} \right\}$ and $\forall s \in \left\{ {1,2, \cdots ,{N_{\rm{L}}}} \right\}$.

\subsubsection{The scale of the SAoS}\label{sec:3-2-1}
We know that $M_{\rm L}$ and $N_{\rm L}$ determine the scale of the SAoS. According to \eqref{eq:S}, the mean receive signal power $S$ contains four components. When the VR of RIS tiles is omnidirectional, the first component, derived for the deterministic LoS channel between the BS/MS and the RIS tiles, increases with the total number of the RIS elements $M_{\rm L}N_{\rm L}$ in the order of $M^2_{\rm L} N^2_{\rm L}$. On the other hand, the last three components of $S$, resulting from the existent of at least one NLoS channel between the BS/MS and the RIS tiles, increase with $M_{\rm L}N_{\rm L}$ in the order of $M_{\rm L} N_{\rm L}$. This implies that when the LoS channel between the BS/MS and the RIS pathes become stronger, the performance gain achieved by the increment of the RIS element number will be more significant.
\subsubsection{The deployment of the RIS tiles}

The effect of RIS tiles arrangement can be seen from the free space attenuation. This effect is larger for the MS than that of the BS, since the MS is much closer to the SAoS compared with the BS and it may not located in the VR of some RIS pathes. As we can be see from the first two component of $S$, putting all RIS pathes closer to the MS so that all VRs cover the same MS will increase $\alpha_{n,c}$ ( $\forall n \in \left\{ {1,2, \cdots ,{M_{\rm{L}}}} \right\}$ and $\forall c \in \left\{ {1,2, \cdots ,{N_{\rm{L}}}} \right\}$ ), and thus can improve the mean receive signal power. This is what the traditional architecture that designs the RIS as a tight uniform panel, achieves. In practical, however, the MS is actually randomly distributed in the blind coverage area. In this case, the traditional RIS architecture may not provide enough SNR for the MS in the edge of blind coverage area due to the limited VR of RIS. On the contrary, the SAoS, covering more larger area than its counterpart in traditional RIS architecture, can provide even services to MS in different positions of the blind coverage area. Therefore, under the same scale of RIS elements, designing the total VR of SAoS proportional to blind coverage area instead of being a compact one will promote the ergodic spectral efficiency for the MS distributed in different positions.\par


\section{Numerical results}\label{sec:5}
In this section, the tightness of the approximation of ergodic spectral efficiency is firstly verified, then effects of the SAoS design on the ergodic spectral efficiency are also evaluated. The system carrier frequency $f_c$ is set as 28GHz and the transmit SNR is $\frac{1}{\sigma^2}$. Without loss of generality, $K_{\rm B}$ and $K_{\rm M}$ are assumed to be the same and denoted by $K$. The SAoS is deployed on the $yoz$ plane and its center is assumed to be ${\bf p}_{\rm S}=(0,0,5)$ in the Cartesian coordinates. We fix the vertical size of SAoS as $M_{\rm{L,v}}=3$ and $d_{\rm{L,v}}=1$. The vertical and horizontal element space in each RIS tile are both set as $\frac{\lambda}{6} $, where $\lambda$ is the carrier wavelength. The VR of RIS tile on the $xoy$ plane is a $90^\circ$ section and it is omnidirectional in the $z$ axis. The BS is assumed to be equipped with UPA of size $4\times 4$ and is located at ${\bf p}_{\rm B} = (100,-100,10)$. The blind coverage area of the BS is assumed to be a street canyon. The MS is randomly distributed in that blind coverage area with position ${\bf p}_{\rm M} = (x_{\rm M},y_{\rm M},z_{\rm M})$ where $x_{\rm M}\in [3,5]$, $y_{\rm M}\in [-9,9]$, $z_{\rm M}\in [-2,2]$.\par

\begin{figure}[!t]
\centering
    \includegraphics[width=0.5\textwidth]{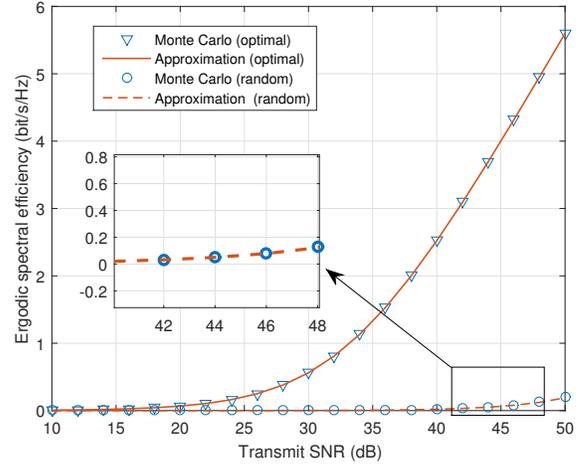}
\caption{Comparison of Monte Carlo results and ergodic spectral efficiency approximations when the reflection coefficients are optimally and randomly designed, respectively.}
\label{Fig.design}
\vspace{-0.7cm}
\end{figure}

The tightness of the approximated ergodic spectral efficiency in \textit{Proposition} \ref{pro:1} is illustrated in Fig.\ref{Fig.design}. Here the horizontal size of SAoS is set as $M_{\rm {L,h}}=8$ and $d_{\rm {L,h}}=3$. Each RIS tile contains $3\times 3$ elements. It can be seen from Fig.\ref{Fig.design} that the approximation matches well with the Monte Carlo results, which proves the correctness of \textit{Proposition} \ref{pro:1}. Furthermore, by optimally designing the reflection coefficients according to \eqref{eq:reflect-design}, the ergodic spectral efficiency is significant higer than that of using random reflection coefficients. As the transmit SNR increases, this performance gain can be further enlarged. This reveals that properly designing the reflection coefficients is the key to exploit the potentialities of the RIS in wireless communication systems. Therefore, in the following simulation, the reflection coefficients are all optimally designed.\par

\par

\begin{figure}[!t]
\centering
    \includegraphics[width=0.47\textwidth]{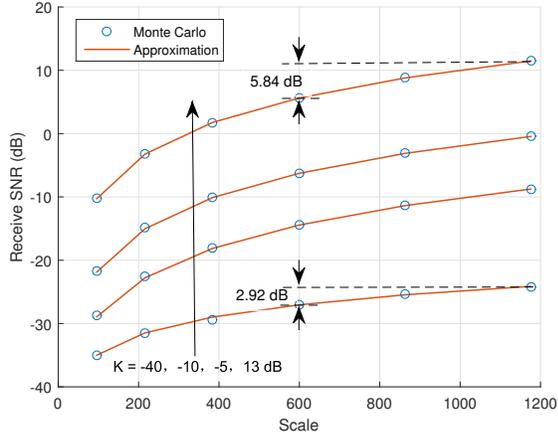}
\caption{Received SNR versus the scale of SAoS.}
\label{Fig.scale}
\vspace{-0.5cm}
\end{figure}

Next, we evaluate the received SNR versus the scale of SAoS under different Rician factor as shown in Fig. \ref{Fig.scale}. The number of RIS tiles remains unchanged to that of Fig. \ref{Fig.design} while the scale of each RIS tile increases from $2\times2$ to $7\times7$. As shown in Fig. \ref{Fig.scale}, under the same scale, increasing the Rician factor can bring additional received SNR gain, which shows the importance of deploying the SAoS in the LoS of the BS and MS. In additional, it is worth noting that the received SNR grows with the square of the SAoS scale under a lager Rician factor, while the rise is linear under small Rician factor. For example, when the scale of SAoS increases from $600$ to $1176$, nearly doubled, a $5.84$ dB gain is obtained under $K=13$ dB while the gain is only $2.92$ dB under $K=-40$ dB. This scaling law is identical to our analysis in Section \ref{sec:3-2-1}, since the mean receive signal power is dominated by the first term and the last three terms of \eqref{eq:S} under lager and small Rician factor, respectively.
\par
\begin{figure}[!t]
\centering
    \includegraphics[width=0.47\textwidth]{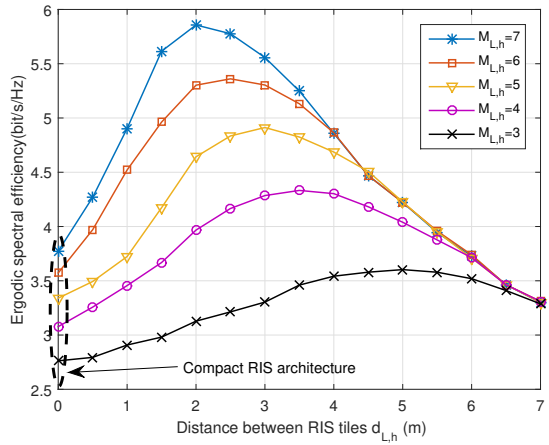}
\caption{Ergodic spectral efficiency versus horizontal distance of RIS tiles for  different horizontal RIS tiles number.}
\label{Fig.distance}
\vspace{-0.8cm}
\end{figure}\par
Finally, the effect of RIS tiles deployment is investigated. Fig. \ref{Fig.distance} demonstrates the ergodic spectral efficiency versus the horizontal deployed distance between RIS tiles. The ergodic spectral efficiency is averaged over any possible position of the MS within the blind coverage area. We keep the size of each RIS tile as $3\times3$ and increase the horizontal number of RIS tiles from $3$ to $7$. As shown in Fig. \ref{Fig.distance}, when $M_{\rm {L,h}}$ is fixed, the ergodic spectral efficiency firstly increases with $d_{\rm {L,h}}$, and then decreases after reaching the peak. Therefore, there exists an optimal deployed distance for the SAoS to maximize the ergodic spectral efficiency. The optimal deployed distance will make the VR of SAoS approximate to the blind coverage area. For example, as we can see from Fig. \ref{Fig.distance}, the optimal deployed distance is $2$ m when $M_{\rm {L,h}}=7$. We know that at $x=3$ each RIS tile can cover $6$ m in $y$ axis with a $90^\circ$ VR section, thus $7$ RIS tiles spacing $2$ m can cover $18$ m, which happens to be the length of blind coverage area in $y$ axis. Fig. \ref{Fig.distance} also demonstrate that when $M_{\rm {L,h}}$ decreases, the optimal horizontal distance will be larger since the total VR of SAoS should keep unchanged. Compared with the compact RIS architecture, which has a zero horizontal distance, the SAoS obtains a higher ergodic spectral efficiency when deployed properly. In additional, it should be noted that when $d_{\rm {L,h}}$ keep increasing, the VR of some RIS tiles will be out of the blind coverage area, resulting in the convergence of ergodic spectral efficiency for different $M_{\rm {L,h}}$.



\section{Conclusion}\label{sec:6}
This paper has investigated the RIS aided anti-blockage mmWave communication systems. To deal with the blockage problem, a SAoS architecture for RIS was proposed. We firstly derived an approximation for the ergodic spectral efficiency. Based on the approximated ergodic spectral efficiency, the optimal reflection coefficients of RIS can be obtained. Furthermore, we found that the received SNR at the BS increases quadratically to the number of the RIS under large Rician factor, while this scaling law is linear under small Rician factor. Considering the VR of RIS tiles, the RIS should be sparsely deployed to cover a lager blind coverage area. Numerical results verified the tightness of the approximated ergodic spectral efficiency and revealed the significant performance gain obtained by optimally designing the reflection coefficients. The scaling law for received SNR and the optimal distance between RIS tiles were also illustrated in numerical results.

\appendices

\section*{\sc \uppercase\expandafter{\romannumeral7}. Acknowledgement}
This work was supported in part by the National Key Research and Development Program 2018YFA0701602, the National Science Foundation of China (NSFC) for Distinguished Young Scholars with Grant 61625106, and the NSFC under Grant 61941104.

\begin{small}

\end{small}
\end{document}